\documentclass{PoS}
\usepackage{graphicx}
\usepackage{wrapfig}
\usepackage[fleqn]{amsmath}
\usepackage{amssymb}
\newcommand{\be}{\begin{equation}}
\newcommand{\ee}{\end{equation}}
\newcommand{\bea}{\begin{eqnarray}}
\newcommand{\eea}{\end{eqnarray}}
\newcommand{\F}{\phantom {1}}

\title{%
\begin{picture}(0,0)(0,0)%
\put(0,75){\makebox(0,0)[l]{\textnormal{\normalsize DESY 10-207}}}%
\put(0,60){\makebox(0,0)[l]{\textnormal{\normalsize HU-EP-10/63}}}%
\end{picture}%
Momentum dependence of the topological susceptibility with overlap fermions}

\ShortTitle{Momentum dependence of the topological susceptibility with overlap fermions}

\author{\speaker{Yoshiaki Koma}$^{a}$, 
Ernst-Michael~Ilgenfritz$^{b}$, Karl~Koller$^{c}$, Miho Koma$^{a}$, Gerrit~Schierholz$^{d}$,
Thomas~Streuer$^{e}$, Volker~Weinberg$^{f}$\\
~\\
$^{a}$ Numazu College of Technology, Numazu, Shizuoka 410-8501, Japan\\
$^{b}$ Institut f\"ur Physik, Humboldt-Universit\"at zu Berlin, 12489 Berlin, Germany\\
$^{c}$ Fakult\"at f\"ur Physik, Ludwig-Maximilians-Universit\"at M\"unchen, 
80333 M\"unchen, Germany\\
$^{d}$ Deutsches Elektronen-Synchrotron DESY, 22603 Hamburg, Germany\\
$^{e}$ Institut f\"ur Theoretische Physik, Universit\"at Regensburg,
93040 Regensburg, Germany\\
$^{f}$ Leibniz-Rechenzentrum der Bayerischen Akademie der Wissenschaften,
85748 Garching bei M\"unchen, Germany}

\abstract{
Knowledge of
the derivative of the topological susceptibility at zero momentum
is important for assessing the validity of the Witten-Veneziano formula for 
the \( \eta' \) mass, and likewise for the resolution of the EMC proton spin 
problem. We investigate the momentum dependence of the topological 
susceptibility and its derivative at zero momentum 
using overlap fermions in quenched lattice QCD simulations.
We expose the role of the low-lying Dirac eigenmodes for the topological 
charge density, and find 
a negative value for the derivative. 
While the sign of the derivative is consistent with the QCD sum rule
for pure Yang-Mills theory, the absolute value is overestimated 
if the contribution from higher eigenmodes is ignored.
}

\FullConference{The XXVIII International Symposium on Lattice Filed Theory\\
June 14-19,2010\\
Villasimius, Sardinia Italy}

\begin{document}

\section{Introduction}

The derivative at zero momentum of the topological susceptibility 
$\chi(k^2)$ in momentum space, $\chi^{\prime}(0)$, is important to control
the validity of the Witten-Veneziano formula for the \( \eta' \) 
mass~\cite{Witten:1979vv,Veneziano:1979ec}. 
The derivation of this formula 
presupposes that the momentum dependence of the topological susceptibility 
$\chi(k^{2})$ 
is moderate from $k^{2}=0$ to $m_{\eta'}^{2}$, 
i.e. $|\chi^{\prime} m^2_{\eta^{\prime}}|\ll \chi$. Moreover, the large-$N_c$ limit is
implied.
The derivative is also helpful to analyze the EMC
proton spin problem~\cite{Ashman:1987hv}
(see \cite{Shore:2007yn} for a review). 
Therefore it is of great interest to investigate $\chi^{\prime}(0)$ starting 
from QCD.

\par
Based on various kinds of QCD sum rules, the following estimates 
of $\chi^{\prime}(0)$ have been reported.
For pure gauge (Yang Mills) theory,
$\chi^{\prime}(0)|_{\rm YM}=-(7\pm 3~{\rm MeV})^{2}$~\cite{Narison:1990cz}, 
and with account of dynamical quarks,
$\sqrt{\chi^{\prime}(0)} \approx 23.2 \pm 2.4$~MeV~\cite{Narison:1994hv}, 
$\chi^{\prime}(0)=(2.3 \pm 0.6)\times 10^{-3}$~GeV$^{2}$~\cite{Ioffe:1998sa}, 
$\sqrt{\chi^{\prime}(0)}=26.4 \pm 4.1$~MeV 
($\chi^{\prime} \approx 0.7\times 10^{-3}$~GeV$^{2}$)~\cite{Narison:1998aq}, 
and $\chi^{\prime}(0)\approx 1.82\times 10^{-3}$~GeV$^{2}$~\cite{Pasupathy:2005ag}.

\par
On the other hand, there are few lattice QCD simulations,
which resulted in the following values. 
In quenched lattice theory,
$\chi^{\prime}(0) |_{\rm SU(2)}= -(9.84 \pm 0.91~{\rm MeV})^{2}$~\cite{Briganti:1990fj}, 
$\chi^{\prime}(0) |_{\rm SU(3)}= -(13 \pm 16~{\rm MeV})^{2}$~\cite{DiGiacomo:1992wg}, 
and
full lattice QCD with staggered fermions,
$\sqrt{\chi^{\prime}(0)}=19\pm 4~{\rm MeV}$~\cite{Boyd:1997nt}. 
In these simulations, the topological charge density has been
defined using the (unimproved) field strength tensor constructed in terms 
of plaquette variables, and the cooling method has been applied to eliminate 
the ultraviolet (UV) noise.
Here, one may notice that the sign of the \( \chi^{\prime}(0)\) estimated by 
the QCD sum rules is negative in pure Yang-Mills theory while it is 
positive in full QCD,
and the lattice results seem to support this tendency within numerical errors.

\par
In this paper, we are going to use a different approach in order to further 
investigate the derivative of the  topological susceptibility at zero momentum
using lattice QCD simulations.
We shall use overlap fermions~\cite{Neuberger:1997fp,Neuberger:1998wv} 
to define the topological charge density~\cite{Niedermayer:1998bi}. 
This approach is extremely useful to clarify the topological structure of 
the QCD vacuum, since it preserves exact chiral symmetry on the 
lattice~\cite{Luscher:1998pqa} and satisfies the index 
theorem~\cite{Hasenfratz:1998ri}. In this approach, it is also possible to 
expose the role of the low-lying Dirac eigenmodes for the topological 
structure of the QCD vacuum~\cite{Ilgenfritz:2007xu}.

\section{Topological charge density and susceptibility from overlap fermions}

\par
In order to define the topological charge density $q(x)$ on the 
lattice with the lattice spacing \( a \), we employ the massless 
overlap Dirac  operator~\cite{Neuberger:1997fp,Neuberger:1998wv} 
defined by
\bea
D =\frac{\rho}{a}(1+\frac{X}{\sqrt{X^{\dagger}X}}) \; ,
\quad
X=D_{W}-\frac{\rho}{a}\; ,
\label{eqn:overlap_operator}
\eea
where $D_{W}$ is the Wilson-Dirac operator. 
We set $\rho=1.4$, a value
identified for the lattices in use as an optimal choice.
Details of our implementation are described in~\cite{Ilgenfritz:2007xu}.
Overlap fermions possess exact chiral symmetry on the
lattice~\cite{Luscher:1998pqa}
and provide $n_{-} + n_{+}$ exact zero modes, $D \psi_{n}^{\pm}=0$,
with $n_{-}$ ($n_{+}$) being the number of modes with negative (positive) 
chirality:
$\gamma_{5} \psi_{n}^{-}= -\psi_{n}^{-}$
 and $\gamma_{5} \psi_{n}^{+}= +\psi_{n}^{+}$.
The index is given by $Q = n_{-} - n_{+}$~\cite{Hasenfratz:1998ri}.
The non-zero modes with eigenvalue $\lambda$,
$D \psi_{\lambda} = \lambda \psi_{\lambda}$,
occur in complex conjugate pairs $\lambda$ and $\lambda^{*}$ and satisfy
$\sum_{x}^{}p_{\lambda 5}(x)= 
\sum_{x}^{} {\rm Tr}~\psi_{\lambda}^{\dagger} (x)
 \gamma_{5}\psi_{\lambda}(x)=0$,
where Tr should be regarded as the sum over color and spinor indices.

\par
Then, the topological charge density $q(x)$ is given 
by~\cite{Niedermayer:1998bi}
\bea
q(x) \equiv
- \mbox{Tr} \left [ \gamma_{5}  (1  -\frac{a}{2} D(x,x) )\right ] \; ,
\label{eqn:q}
\eea
which satisfies the index theorem with 
$Q= \sum_{x} q(x)  \in \mathbb{Z}$.
The contribution of the low-lying Dirac eigenmodes to $q(x)$
can be exposed by applying the eigenmode 
expansion~\cite{Horvath:2002yn,Koma:2005sw,Weinberg:2006ju},
\bea
q(x;\lambda_{\rm cut}) = -
\sum_{|\lambda|\leq \lambda_{\rm cut}}
\left( 1-\frac{\lambda}{2} \right ) p_{\lambda 5}(x) \; , 
\label{eqn:q-expansion}
\eea
where the cut-off $\lambda_{\rm cut}$ implies
a kind of UV filtering, and
$p_{\lambda 5}(x)= \psi_{\lambda}^{\dagger}(x)
\gamma_{5}\psi_{\lambda}(x)$
is the local chirality of the mode corresponding to the eigenvalue $\lambda$.
Note that the UV filtering maintains the index theorem independently of the 
cut-off such that $Q=\sum_{x}q(x;\lambda_{\rm cut})$. This is because the index 
is computed only from the zero modes. 
Actually, it is observed that all zero modes of a lattice configuration
occur with the same chirality.
Truncating the expansion at $\lambda_{\rm cut}$ acts like an 
UV filter by removing certain short distance fluctuations from the local
density $q(x)$.

\par
In the continuum limit, the momentum dependent topological susceptibility is 
defined as
\bea
\chi (k^{2})
=\int d^{4}x ~e^{ikx} C_{q}(x)\;,
\label{eqn:sus1}
\eea
where $C_{q}(x) = \langle T(q(x) q(0) )\rangle$ is the two-point correlation 
function of the
topological charge density $q(x)$.
Eq.~\eqref{eqn:sus1} can be expanded with respect to $k^{2}$ 
and one obtains the expression of the derivative of the topological 
susceptibility at zero momentum
\bea
\chi^{\prime} (0)=\frac{d\chi (k^{2})}{d k^{2}} \Biggr |_{k^{2}=0}
=-\frac{1}{8}\int_{}^{}d^{4}x ~x^{2}C_{q}(x)\; .
\label{eqn:sus1-derivative}
\eea
The lattice version of Eqs.~\eqref{eqn:sus1} 
and~\eqref{eqn:sus1-derivative} reads
\bea
\chi (\hat{k}^{2}) =  \sum_{x \in V} e^{ikx} C_{q}(x) \; ,
\label{eqn:sus2-fourier} 
\eea
\bea
\chi^{\prime} (0) =  -\frac{1}{8}\sum_{x\in V} x^{2} C_{q}(x) \; ,
\label{eqn:sus2-derivative} 
\eea
respectively, where
\bea
C_{q}(r)=\frac{1}{V}\sum_{x\in V} \langle q(x+r) q(x) \rangle  \;.
\label{eqn:qqcor}
\eea
As we consider the lattice volume $V=L^{3}T$ and
impose periodic boundary conditions in all directions,
the momentum is discrete as
\bea
\hat{k}_{\mu}
=\frac{2}{a}\sin(\frac{ak_{\mu}}{2})\;,\qquad
k_{\mu}
=
2\pi  (\frac{n_{1}}{L}, \frac{n_{2}}{L}, \frac{n_{3}}{L}, \frac{n_{4}}{T})\;,
\label{eqn:momentum}
\eea
with $n_{\mu}$ being integer; $(n_{1},n_{2},n_{3})$ runs from 0 to $(L/a)-1$  
and $n_{4}$ from 0 to $(T/a)-1$.

\section{Numerical results}

\begin{table}[!t]
\centering
\caption{Details of the ensemble used in this study.
The L\"uscher-Weisz gauge action is used, 
$\beta$ denotes the inverse coupling squared, 
$a$ the lattice spacing (determined from the pion decay constant), 
and $V=L^{3}T$ the lattice volume. $\chi(0)$ is the topological 
susceptibility.}~\\[-0.2cm]
\begin{tabular}{|c|c|cc|c|l|}
\hline
$\beta$ & $a$ [fm]  & $(L/a)^{3}(T/a)$  & (fm$^{4}$)
& \# of config. 
& \( \chi (0) \) ~[GeV\( ^{4} \)]\\
\hline
8.45 & 0.105  &  $16^{3}32$  & (15.9)  &267  
&\( 8.2(7)\F  \times 10^{-4} \)
\\
 \hline
  8.45 & 0.105  & $12^{3}24$&(5.0)\F &  116 
  &   \( 9.1(13)  \times 10^{-4} \)
  \\
\hline
 8.10 & 0.142 & $12^{3}24$  &(16.9)  &  254 
 &\( 8.6(8)\F  \times 10^{-4} \)
 \\
\hline
8.00 & 0.157 &$16^{3}32$  &(79.6)  & 444 
& \( 9.1(6)\F  \times 10^{-4}  \)
\\
\hline
\end{tabular}
\label{tab:tab1}
\end{table}

\par
We used several ensembles of zero-temperature quenched configurations 
generated by means of the L\"uscher-Weisz gauge action~\cite{Galletly:2003vf}
(see Table~\ref{tab:tab1}).
This action is suitable for topological studies since
dislocations are greatly suppressed.

\par
\begin{figure}[!b]
\centering
\includegraphics[width=7.5cm]{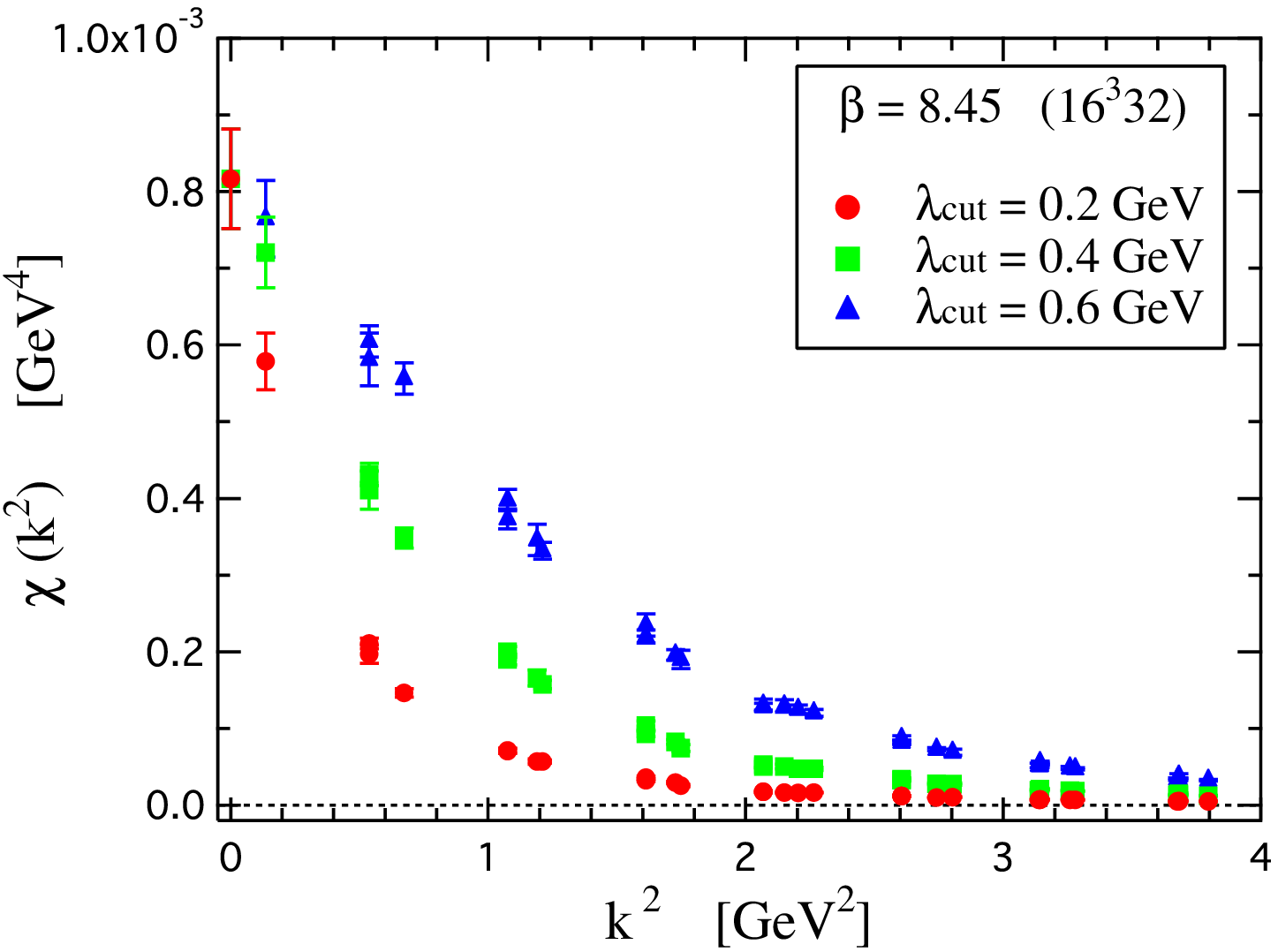}
\includegraphics[width=7.5cm]{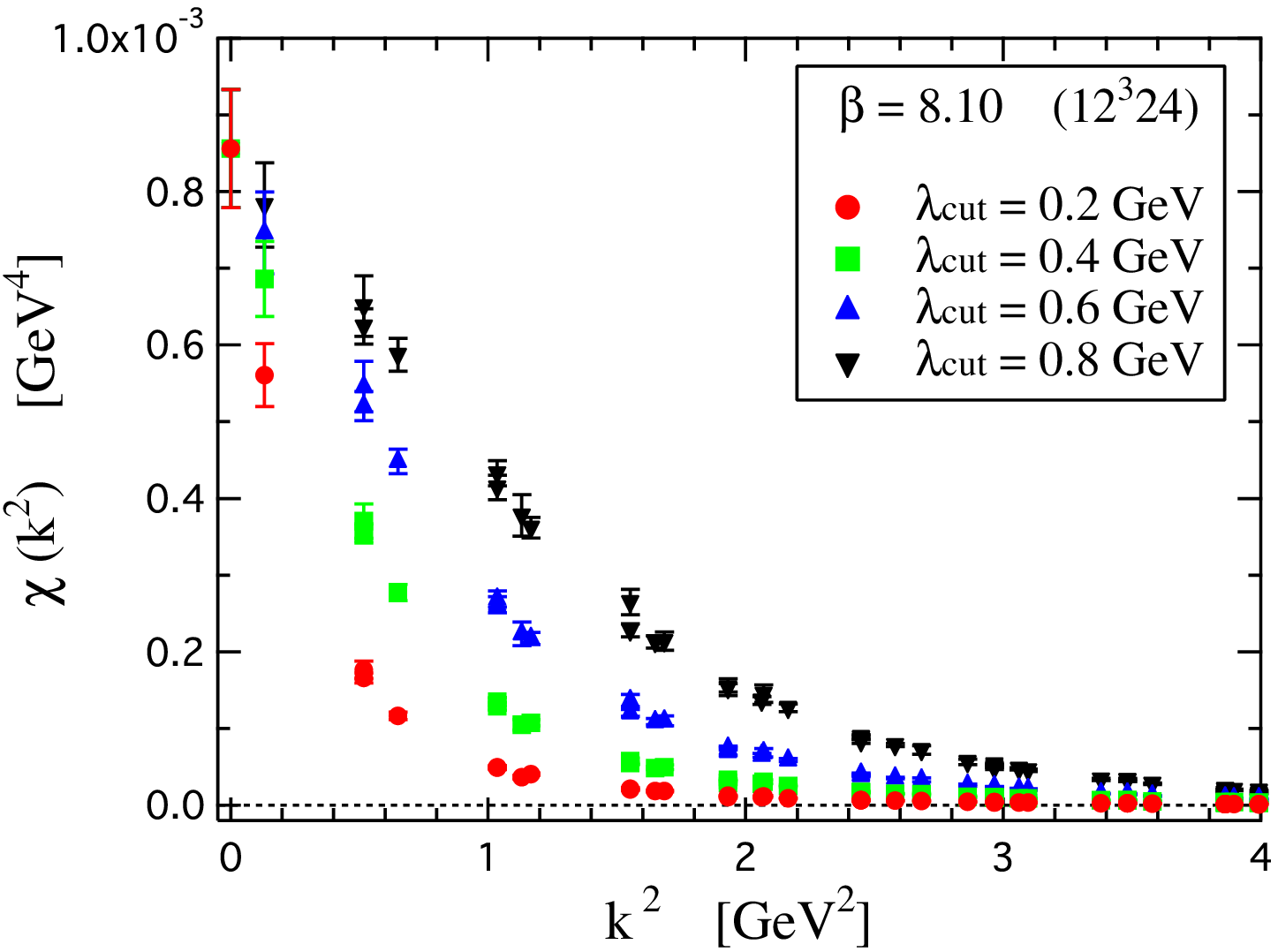}
\caption{The momentum dependence of the
topological susceptibility at 
$\beta=8.45$ on the $16^{3}32$ lattice (left).
There are two data points at some momenta, 
which are due to the violation of rotational symmetry,
and are expected to converge to one point in the continuum limit.
The same as the left figure,
but at $\beta=8.10$ on the $12^{3}24$ lattice (right) .
} 
\label{fig:fig1}
\end{figure}

We plot $\chi(\hat{k}^{2})$ as measured at \( \beta=8.45 \) on 
the \( 16^{3}32 \) lattice in Fig~\ref{fig:fig1} (left), and as measured 
at \( \beta=8.10 \) on the \( 12^{3}24 \) lattice in Fig.~\ref{fig:fig1} (right),
for various cut-off values of the (imaginary part of the) eigenvalue, 
\( \lambda_{\rm cut} = 0.2 ~-~ 0.8 ~{\rm GeV}\).
In these analyses, the eigenvalue $\lambda$ of Eq.(\ref{eqn:overlap_operator})
is always replaced by
\( \lambda_{\rm imp} = (1-a\lambda/2\rho)^{-1}\lambda \). These are the eigenvalues
of the improved massless overlap operator $D_{\rm imp}$~\cite{Galletly:2003vf}. 
The improvement projects the eigenvalues of 
$D$ stereographically onto the imaginary axis. We find that the largest value 
of \( \chi (\hat{k}^{2}) \) is at \( \hat{k}^{2}=0 \) for each cut-off value 
of the eigenvalues \( \lambda_{\rm cut} = 0.2 ~-~ 0.8 ~{\rm GeV}\) and that 
the functions \( \chi (\hat{k}^{2}) \) are monotonously decreasing as functions
of the momentum for all cut-off values.
The qualitative behavior of \( \chi (\hat{k}^{2}) \) does not depend on the 
lattice spacing.

\par
\begin{wrapfigure}{r}{8cm}
\vspace*{-0.4cm}
\centering
\includegraphics[width=7.5cm]{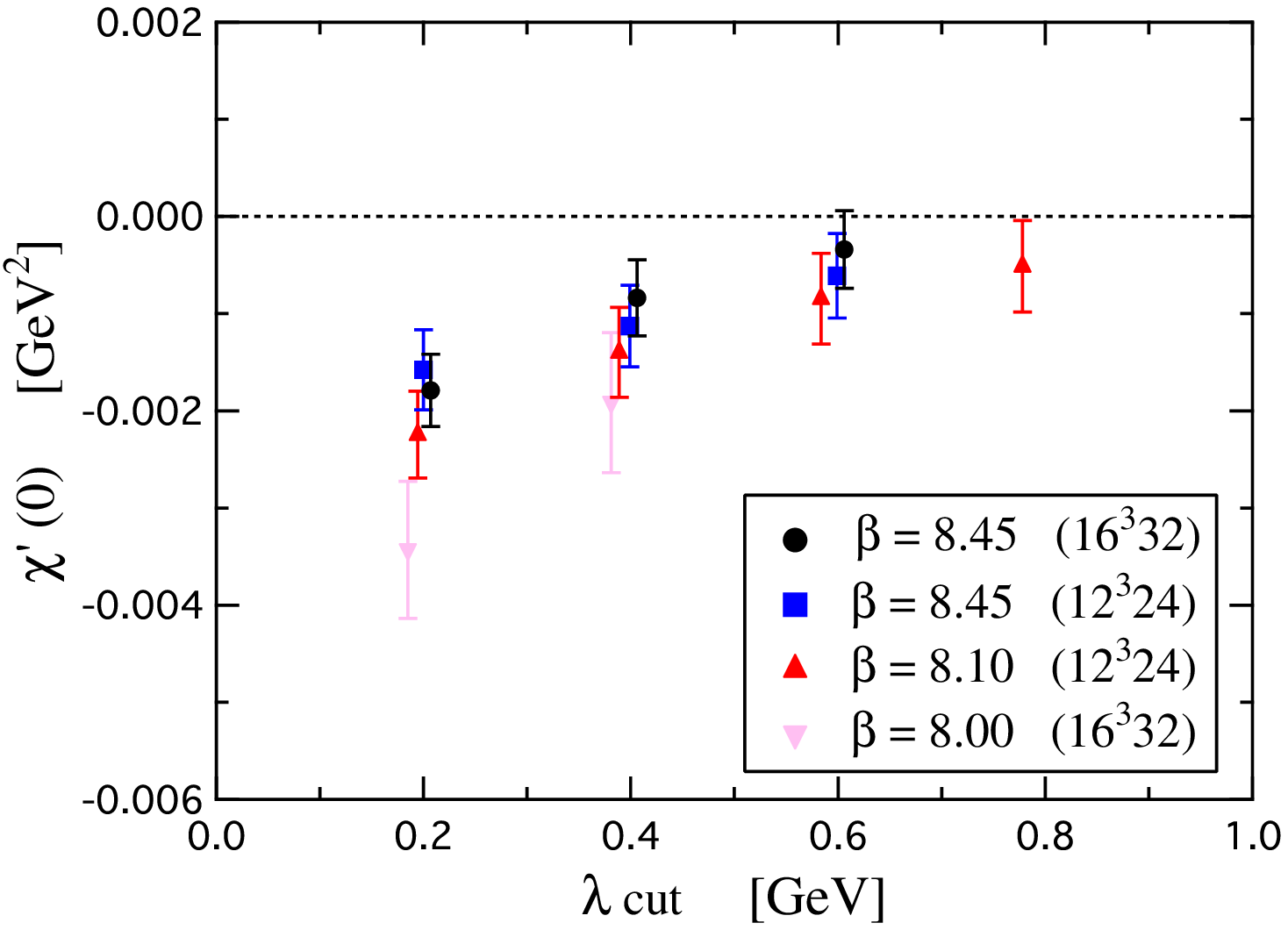}
\caption{
The slope $\chi^{\prime}(0)$ as a function of the cut-off 
$\lambda_{\rm cut}$.} 
\label{fig:fig3}
\includegraphics[width=7.5cm]{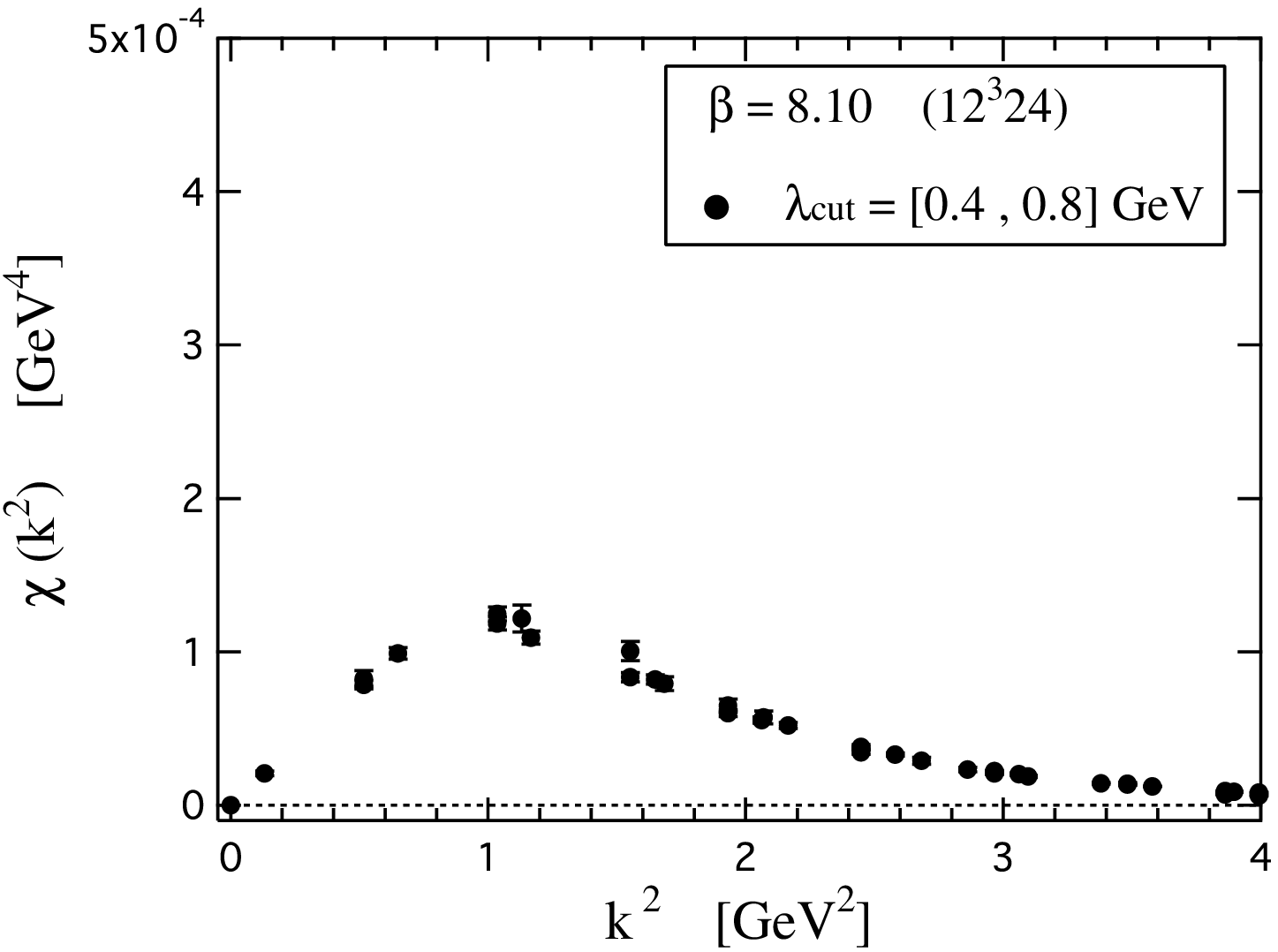}
\caption{The same as in Fig.~1, but for the ensemble 
with $\beta=8.10$ on the $12^{3}24$ lattice, when the zero modes
and the lowest eigenmodes are removed from the evaluation of
\( \chi(\hat{k}^{2}) \). Only eigenmodes between 
\( 0.4 \leq \lambda \leq 0.8~{\rm GeV} \) are taken into account.} 
\label{fig:fig5}
\end{wrapfigure}

The result of the \( \chi^{\prime}(0) \) 
according to Eq.~\eqref{eqn:sus2-derivative} is plotted in Fig.~\ref{fig:fig3}.
We do not see a significant volume dependence at \( \beta=8.45 \) 
for which two lattice volumes are available. Since the physical volume of 
the \( 12^{3}24 \) lattice at \( \beta=8.45 \) is the smallest among our
lattice ensembles (see Table~\ref{tab:tab1}), we expect that the finite 
volume effect will not spoil our observation.
The physical scales of the cut-off values at different \( \beta \) values
are chosen approximately corresponding to each other, so as to examine the
lattice spacing dependence of \( \chi^{\prime}(0) \).
In the result we find a reasonable scaling behavior of \( \chi^{\prime}(0) \) 
at corresponding cut-off values \( \lambda_{\rm cut} \),
apart from the data of the coarsest lattice at \( \beta=8.00 \).

\par
In any case, we find an interesting behavior:
as the cut-off increases, \( | \chi^{\prime}(0) |\) becomes smaller.
In other words, \( | \chi^{\prime}(0) |\) takes on the largest values
if only the lowest eigenmodes are included in the fluctuations of the
topological density.
If one, considering the Witten-Veneziano formula,
ignores the necessary condition of 
\( |\chi^{\prime}(0) m_{\eta'}^{2}| \ll \chi (0)\), 
it seems that the topological susceptibility at zero momentum \( \chi (0) \)
is the only important parameter for the \( \eta' \) mass.
Since \( \chi (0) \) depends only on the (number and chirality of the )
zero modes, one might erroneously conclude that the zero modes 
are sufficient to describe the topological structure of the 
QCD vacuum. However, the behavior of \( \chi^{\prime}(0) \)
indicates that fluctuation of the topological charge density at all scales 
are necessary to warrant that 
\( |\chi^{\prime} (0)m_{\eta'}^{2}|  \ll  \chi (0)\)
can be satisfied.
\\[-0.6cm]
For instance, taking a typical value at \( \beta=8.10 \),
we have \( \chi^{\prime}(0) \approx -0.001~{\rm GeV}^{2} \)
($\chi^{\prime}(0) \approx - (32~{\rm MeV})^2$)
with \( \chi (0) \approx 8.6\times 10^{-4}~{\rm GeV}^{4} \),
we estimate \( \chi (0)/|\chi^{\prime}(0) | \approx 0.86~{\rm GeV^{2}}
=(930~{\rm MeV})^{2}\), which
is comparable with \( m_{\eta'} = 958~{\rm MeV}\).
This in turn means that not only the lowest eigenmodes (not to mention
only the zero modes) but also 
the higher eigenmodes have to play an important role for the topological structure 
of the QCD vacuum.

\par
It is then quite intriguing to investigate which value
the derivative \( \chi^{\prime}(0) \) approaches if more and more contributions
from higher eigenmodes (possibly all eigenmodes) are included.
We investigate this for the case of \( \beta=8.10 \) on the \( 12^{3}24 \)
lattice based on the definition of the full topological charge density in 
Eq.~\eqref{eqn:q}, which was evaluated without mode expansion.
We find the value of the 
derivative \( \chi^{\prime}(0) = -2.4(16) \times 10^{-3}~{\rm GeV^{2}} \).
However, we note that the number of configurations used in this analysis is 
only 53~\cite{Ilgenfritz:2007xu}
and that the topological susceptibility is overestimated for this subsample as
\( \chi(0)= 1.3(3) \times 10^{-3}~{\rm GeV^{4}} \). 
The limited number of 
The small number of configurations is dictated by the high cost
of evaluating Eq.~\eqref{eqn:q} using the overlap Dirac operator.
Therefore, the value of 
\( \chi^{\prime}(0) \) based on the all-scale topological density (formally
\( \lambda_{\rm cut} \to \infty \) ) cannot be considered as conclusive.

\par
Next we examine the role of the zero and lowest modes
for \( \chi (\hat{k}^{2}) \).
In Fig.~\ref{fig:fig5}, we plot the \( \chi (\hat{k}^{2}) \)
for the ensemble with \( \beta=8.10 \) on the \( 12^{3}24 \) lattice, 
where only the eigenmodes between \( 0.4 \leq \lambda \leq 0.8~{\rm GeV} \)
are used to define the topological charge 
density, i.e., the zero modes and very low-lying modes are removed.
We find that the slope at \( \hat{k}^{2}=0 \) turns to be positive.
\( \chi (\hat{k}^{2}) \) increases from zero at \( \hat{k}^{2} = 0 \)
(due to the subtraction of zero modes from the topological density)
to a certain maximum value (near to \( \hat{k}^{2} \approx 1~{\rm GeV}^2 \)  )
and then decreases.

This leads us to the speculation
that -- if the zero modes are dynamically suppressed 
and the relative weight of sectors with smaller $|Q|$ will increase 
in the average --
the sign of the derivative \( \chi^{\prime}(0) \) may change from negative 
to positive. This may be the case in full QCD in the chiral limit.

\section{Summary}

\par
Using the overlap fermion formalism, we have investigated the momentum 
dependence of the topological susceptibility and 
its slope at zero momentum, \( \chi^{\prime}(0) \).
Overlap fermions preserve exact chiral symmetry on the lattice and
possess exact zero modes, which allow us to unambiguously compute the
index $Q$ of vacuum configurations.
We have found that \( \chi^{\prime}(0) \) depends on the number of 
eigenmodes determining the resolution in the definition of the topological 
charge density.
The more the cut-off applied to the eigenvalues of the Dirac operator 
increases, the more the absolute value 
of the slope \( |\chi^{\prime}(0)| \) decreases.
In other words, \( |\chi^{\prime}(0)| \) with only the lowest-lying eigenmodes
included turns out too large to guarantee that the 
technical assumption underlying
the Witten-Veneziano formula is fulfilled.
From this point of view, fluctuation of the topological charge density at all
scales, represented by higher eigenmodes, should also be taken into account. 
Therefore it is not correct to argue that only zero modes and low-lying 
eigenmodes are relevant for the topological structure of the QCD vacuum.

\par
As demonstrated in~\cite{Ilgenfritz:2007xu}, 
the topological charge density possesses global sign coherent structures, 
which get increasingly tangled as more and more eigenmodes are included, 
and the all-scale topological charge density has a lower-dimensional, 
laminar structure, together with a lumpy structure inside the sign coherent 
regions. After all, it is found to possess a multifractal structure.
This complicated structure of the QCD vacuum is responsible for the behavior 
of~\( \chi^{\prime}(0) \).

\par
The numerical calculations have been performed at NIC (J\"ulich) and
HLRN (Berlin) as well as at DESY (Zeuthen). 
We thank all institutes for support.
This work was supported in part by Japan Society for the Promotion of
Science (JSPS) and German Research Foundation (DFG),
Japan-German Joint Research Project 2008-2009.
Y.K. was partially supported by the Ministry of Education, 
Science, Sports and Culture,
Japan, Grant-in-Aid for Encouragement of Young 
Scientists (B), No.20740149.
M.K. was supported by Japan Society for 
the Promotion of Science (JSPS),
Grant-in-Aid for JSPS Fellows (20$\,\cdot\,$40152).


\begin{thebibliography}{10}
\itemsep=0cm

\bibitem{Witten:1979vv}
E.~Witten,
{\em {Current algebra theorems for the {U(1)} Goldstone boson}},
Nucl. Phys. {\bf B156}, 269 (1979).

\bibitem{Veneziano:1979ec}
G.~Veneziano,
{\em {{U(1)} without instantons}},
Nucl. Phys. {\bf B159}, 213 (1979).

\bibitem{Ashman:1987hv}
European Muon, J.~Ashman {\em et~al.},
{\em {A measurement of the spin asymmetry and determination of the structure
  function {$g_1$} in deep inelastic muon proton scattering}},
Phys. Lett. {\bf B206}, 364 (1988).

\bibitem{Shore:2007yn}
G.M. Shore,
{\em {The {U(1)$_A$} anomaly and {QCD} phenomenology}},
Lect. Notes Phys. {\bf 737}, 235 (2008), hep-ph/0701171.

\bibitem{Narison:1990cz}
S.~Narison,
{\em {The slope of the {U(1)} topological charge from gluonia sum rules and
  higher order effects on the pseudoscalar meson masses and mixing angles}},
Phys. Lett. {\bf B255}, 101 (1991).

\bibitem{Narison:1994hv}
S.~Narison, G.M. Shore, and G.~Veneziano,
{\em {Target independence of the {EMC - SMC} effect}},
Nucl. Phys. {\bf B433}, 209 (1995), hep-ph/9404277.

\bibitem{Ioffe:1998sa}
B.L. Ioffe and A.G. Oganesian,
{\em {Proton spin content and {QCD} topological susceptibility}},
Phys. Rev. {\bf D57}, 6590 (1998), hep-ph/9801345.

\bibitem{Narison:1998aq}
S.~Narison, G.M. Shore, and G.~Veneziano,
{\em {Topological charge screening and the 'proton spin' beyond the chiral
  limit}},
Nucl. Phys. {\bf B546}, 235 (1999), hep-ph/9812333.

\bibitem{Pasupathy:2005ag}
J.~Pasupathy, J.P. Singh, R.K. Singh, and A.~Upadhyay,
{\em {The derivative of the topological susceptibility at zero momentum and an
  estimate of eta' mass in the chiral limit}},
Phys. Lett. {\bf B634}, 508 (2006), hep-ph/0509260.

\bibitem{Briganti:1990fj}
G.~Briganti, A.~Di~Giacomo, and H.~Panagopoulos,
{\em {A lattice determination of the slope of the topological susceptibility at
  {$q^2 = 0$}}},
Phys. Lett. {\bf B253}, 427 (1991).

\bibitem{DiGiacomo:1992wg}
A.~Di~Giacomo, E.~Meggiolaro, and H.~Panagopoulos,
{\em {A lattice determination of the slope at {$q^2 = 0$} of the topological
  susceptibility in SU(3) Yang-Mills theory}},
Phys. Lett. {\bf B291}, 147 (1992).

\bibitem{Boyd:1997nt}
G.~Boyd, B.~Alles, M.~D'Elia, and A.~Di~Giacomo,
{\em {Topology in {QCD}}},
(1997), hep-lat/9711025.

\bibitem{Neuberger:1997fp}
H.~Neuberger,
{\em {Exactly massless quarks on the lattice}},
Phys. Lett. {\bf B417}, 141 (1998), hep-lat/9707022.

\bibitem{Neuberger:1998wv}
H.~Neuberger,
{\em {More about exactly massless quarks on the lattice}},
Phys. Lett. {\bf B427}, 353 (1998), hep-lat/9801031.

\bibitem{Niedermayer:1998bi}
F.~Niedermayer,
{\em {Exact chiral symmetry, topological charge and related topics}},
Nucl. Phys. Proc. Suppl. {\bf 73}, 105 (1999), hep-lat/9810026.

\bibitem{Luscher:1998pqa}
M.~L{\"u}scher,
{\em {Exact chiral symmetry on the lattice and the Ginsparg- Wilson relation}},
Phys. Lett. {\bf B428}, 342 (1998), hep-lat/9802011.

\bibitem{Hasenfratz:1998ri}
P.~Hasenfratz, V.~Laliena, and F.~Niedermayer,
{\em {The index theorem in {QCD} with a finite cut-off}},
Phys. Lett. {\bf B427}, 125 (1998), hep-lat/9801021.

\bibitem{Ilgenfritz:2007xu}
E.M. Ilgenfritz {\em et~al.},
{\em {Exploring the structure of the quenched {QCD} vacuum with overlap
  fermions}},
Phys. Rev. {\bf D76}, 034506 (2007), 
[arXiv:0705.0018 (hep-lat)].

\bibitem{Horvath:2002yn}
I.~Horvath {\em et~al.},
{\em {On the local structure of topological charge fluctuations in QCD}},
Phys. Rev. {\bf D67}, 011501 (2003), hep-lat/0203027.

\bibitem{Koma:2005sw}
Y.~Koma {\em et~al.},
{\em {Localization properties of the topological charge density and the low
  lying eigenmodes of overlap fermions}},
PoS {\bf LAT2005}, 300 (2006), hep-lat/0509164.


\bibitem{Weinberg:2006ju}
V.~Weinberg {\em et~al.},
{\em {The QCD vacuum probed by overlap fermions}},
PoS {\bf LAT2006}, 078 (2006), hep-lat/0610087.


\bibitem{Galletly:2003vf}
QCDSF-UKQCD collaboration, D.~Galletly {\em et~al.},
{\em {Quark spectra and light hadron phenomenology from overlap fermions with
  improved gauge field action}},
Nucl. Phys. Proc. Suppl. {\bf 129}, 453 (2004), hep-lat/0310028.


\end{thebibliography}
\end{document}